\documentclass[10pt,twocolumn]{article}
\usepackage[top=2cm, bottom=2cm, left=2cm, right=2cm]{geometry}
\usepackage{times}  %
\usepackage[sort&compress,numbers]{natbib}  %
\usepackage[hyphens]{url}
\usepackage{flushend}
\usepackage[hyphens]{url}  %
\usepackage{graphicx} %
\usepackage{tikzsymbols}
\usepackage{graphicx} 
\usepackage{xcolor}
\usepackage{xspace}
\usepackage{xurl}
\usepackage{wrapfig}
\usepackage{booktabs}
\usepackage{comment}
\usepackage{subfigure}
\usepackage{multirow}
\usepackage{amsmath,amsfonts}
\usepackage{bm}
\usepackage{tabularx} 

\usepackage{sectsty}
\sectionfont{\bfseries\Large\raggedright}

\let\oldbibliography\thebibliography
\renewcommand{\thebibliography}[1]{%
  \oldbibliography{#1}%
  \setlength{\itemsep}{2pt}%
}

\usepackage[hang,flushmargin]{footmisc}

\usepackage[compact]{titlesec}
\titlespacing*{\section}{0pt}{*3}{3pt}
\titlespacing*{\subsection}{0pt}{*2}{2pt}
\usepackage{xspace}

\usepackage[skipabove=5pt,skipbelow=5pt,roundcorner=10pt,framemethod=TikZ]{mdframed}

\makeatletter
\def\url@leostyle{%
  \@ifundefined{selectfont}{\def\UrlFont{}}%
  {\def\UrlFont{}}%
}
\makeatother
\usepackage[hyphenbreaks]{breakurl}

\definecolor{darkgreen}{RGB}{0, 100, 0}
\definecolor{linkcol}{rgb}{0.3,0,0}
\definecolor{citecol}{rgb}{0.3,0,0}
\definecolor{urlcol}{rgb}{0.3,0,0}

\usepackage[bookmarks=true, bookmarksnumbered=true, colorlinks=true, linkcolor=linkcol, citecolor=citecol, urlcolor=urlcol, hypertexnames=true]{hyperref}

\newcommand{\descr}[1]{\smallskip\noindent\textbf{#1}}
\definecolor{DarkGreen}{RGB}{0,100,0} 
\definecolor{Darkblue}{RGB}{0, 0, 160} 

\makeatletter
\def\url@leostyle{%
  \@ifundefined{selectfont}{\def\UrlFont{\small}}%
  {\def\UrlFont{}}%
}
\makeatother
\usepackage{algorithm}
\usepackage{algorithmic}
\usepackage{booktabs}
\usepackage{multirow}
\usepackage{subfigure}
\usepackage{xspace}
\usepackage{mdframed}

\newcommand{\name}{{\textsc{Aletheia}}\xspace}

\usepackage{xcolor}
\usepackage{bm}
\usepackage{comment}
\usepackage{amsmath,amsfonts}

\newcommand{\G}{\mathcal{G}}
\newcommand{\V}{\mathcal{V}}
\newcommand{\E}{\mathcal{E}}

\newif\ifcomment
\commenttrue
\ifcomment
\newcommand{\hs}[1]{{\bf \textcolor{brown}{HS: #1}}}
\newcommand{\ik}[1]{{\bf \textcolor{blue}{IK: #1}}}
\newcommand{\hh}[1]{{\bf \textcolor{yellow}{HH: #1}}}
\newcommand{\rev}[1]{{\textcolor{teal}{#1}}}
\else
\newcommand{\hs}[1]{}
\newcommand{\ik}[1]{}
\newcommand{\hh}[1]{}
\newcommand{\rev}[1]{}
\fi

\definecolor{DarkGreen}{RGB}{0,100,0} 

\mdfdefinestyle{lightgrayframe}{
    backgroundcolor=gray!30, 
    linecolor=gray!50,       
    linewidth=1pt,           
    innertopmargin=5pt,      
    innerbottommargin=5pt,
    innerrightmargin=5pt,
    innerleftmargin=5pt,
    roundcorner=5pt          
}

\begin{document}

\title{\bf {\name}: Combating Social Media Influence Campaigns with \\ Graph Neural Networks }

	\author {
	Mohammad Hammas Saeed\textsuperscript{\rm 1}, 
    Isaiah J. King\textsuperscript{\rm 1}, and
	Howie Huang\textsuperscript{\rm 1} \\[1ex]
%
	\textsuperscript{\rm 1}George Washington University
}

\date{}

\maketitle

\begin{abstract}
Influence campaigns are a growing concern in the online spaces.
Policymakers, moderators and researchers have taken various routes to fight these campaigns and make online systems safer for regular users.
To this end, our paper presents \name, a system that formalizes the detection of malicious accounts (or \textit{troll accounts}) used in such operations and forecasts their behaviors within social media networks.
We analyze influence campaigns on Reddit and X from different countries and highlight that detection pipelines built over a graph-based representation of campaigns using a mix of topological and linguistic features offer improvement over standard interaction and user features.
\name uses state-of-the-art Graph Neural Networks (GNNs) for detecting malicious users that can scale to large networks and achieve a 3.7\% F1-score improvement over standard classification with interaction features in prior work.
Furthermore, \name employs a first temporal link prediction mechanism built for influence campaigns by stacking a GNN over a Recurrent Neural Network (RNN), which can predict future troll interactions towards other trolls and regular users with an average AUC of 96.6\%.
\name predicts troll-to-troll edges (TTE) and troll-to-user edges (TUE), which can help identify regular users being affected by malicious influence efforts.
Overall, our results highlight the importance of utilizing the networked nature of influence operations (i.e., structural information) when predicting and detecting malicious coordinated activity in online spaces.

\end{abstract}

\section{Introduction}

Influence campaigns are coordinated online efforts to alter public opinion, craft malicious narratives, and push agendas in online spaces.
Extensive prior research highlights influence operations and their activities on various social media platforms, e.g., Twitter (now X) and Reddit~\cite{xia2021disinformation, zannettou2020characterizing, zannettou2019disinformation, zannettou2019let}.
However, as it stands, malicious influence operations still remains a pertinent issue in cyberspace~\cite{foreign2024disinfo, voa2024crackdown, state2024disinfo}.
These operations are often carried out through dedicated accounts (also referred to as \emph{troll accounts}).
Troll accounts operate by intermingling with real users on a platform, often replying and operating in groups together to simulate a real conversation~\cite{mueller2019mueller, starbird2019collab}.
This aspect of imitating regular behavior makes it difficult to detect troll behavior, since they are distinct from standard social media bots in that they create conversation like real users instead of simply broadcasting template messages, although that might be one of their goals~\cite{saeed2024disinfo}.

\descr{Motivation.}
Prior research shows that troll accounts are \textit{loosely coordinated} actors that rely on pushing each other's messages~\cite{saeed2022troll}, and often maintain online personas to interact with the general public~\cite{linvill2020troll}.
They strategically interject themselves into social media spaces and strategize to spread their agendas, e.g., excessive retweeting in groups and networks of accounts~\cite{stewart2018examining}.
Since these accounts are multi-intent actors, i.e., they reshare information, interact with users and also work together in groups, their reliance on collaborative efforts makes them ideal candidates to be modeled in the form of a network.
Therefore, in this work, we posit the detection of troll accounts and predicting their future behavior as node detection and link prediction tasks in a social media graph.

\descr{Research Problem.}
Although past work highlights the various characteristics of influence campaigns~\cite{wang2023state, ratkiewicz2011detecting, stewart2018examining} and uses some of their characteristics to develop detection systems~\cite{forn2018holistic, alzahbi200mach, saeed2022troll}, it is equally important to develop systems that can predict future troll behavior for preemptive measures given the velocity of information flows in today's online ecosystem. 

In this work, we present \name, a system that models influence campaigns in the form of a network and utilizes a mix of graph and linguistic properties to improve detection performance of malicious spreaders for multiple campaigns on both Reddit and X.
\name consists of a novel framework that stacks a graph neural network (GNN) over a recurrent neural network (RNN) to accurately predict future troll interactions.
The goal is to utilize network information (e.g., real-world social media connections of users) and identify malicious users and behaviors.

\descr{Research Questions.} 
We seek to answer the following research questions.
\begin{itemize}
    \item \descr{RQ1:} Can state-of-the-art graph based approaches improve node classification scores for accounts involved in influence campaigns?
    \item \descr{RQ2:} Can we effectively predict future interactions of a troll account with other troll accounts as well as regular users?
\end{itemize}

To answer the research questions, we study influence campaigns on both Reddit and X from multiple countries.
We use ground truth datasets of accounts released by the platforms.
Our approach is to model the network of social media accounts and identify anomalous coordinated activity. 
\name first models the troll interactions in the form of a graph using sharing patterns (e.g., \emph{replying}, \emph{resharing} and \emph{mentions}).
Next, \name comprises of a multi-modal pipeline built upon GraphSAGE~\cite{hamilton2017inductive} that uses the network graphs to perform node detection, highlighting the efficacy of topological features in combination with language embeddings when representing malicious campaigns.
Our results are consistent across platforms and campaigns showing the importance of structural features in modeling troll activity.
\name also performs a first-of-its-kind temporal link prediction for influence operations over the network and predicts future links with an average AUC of 96.6\%. 
These results are significant since we can use existing snapshots of the network and predict likely users that will be impacted by the troll accounts or when troll accounts interact with one another for amplification of narratives.

\descr{Contributions.}
Overall, we make the following contributions in our work.
\begin{itemize}
    \item We show that modeling influence campaigns as a graph (network) and utilizing topological features (e.g., degree and centrality) to represent nodes in the graph can result in improvement of detection performance for both Reddit and X operations (with an average F1-score of 96.44\% and 97.9\% respectively).
    \item Our results show that developing a multi-modal pipeline, where text embeddings augment a graph-based approach improves detection performance, with a 1.53\% increase in AUC and 2.06\% increase in F1 score from our pipeline built on GraphSAGE.
    \item We present, to the best of our knowledge, the first application of temporal link prediction to influence campaigns where future links can be predicted effectively between \emph{troll-troll} users and \emph{troll-regular} user. \name uses a combination of GraphSAGE and RNN to effectively predict troll interactions with an average AUC of 96.6\%.
\end{itemize}

\descr{Implications.}
Our work paves the way for improved detection and preemptive identification of malicious and manipulative behavior in online spaces.
Developing multi-modal pipelines that can forecast and detect influence efforts in a scalable manner can help curb malicious influence on regular users on social media platforms ensuring user and platform safety.

\descr{Disclaimer.}
We present several case studies that highlight nuanced influence through coordinated operations.
The presented examples might be offensive, harmful and toxic to readers: therefore, reader discretion is advised.

\section{Data}
We use data from both Reddit and Twitter (now X) for our analyses, which are popular social media platforms that have been targeted by large-scale influence operations in the past~\cite{xia2021disinformation, zannettou2020characterizing, zannettou2019disinformation, zannettou2019let}.

\subsection{Reddit}
Reddit is primarily a news aggregation and discussion website where users can subscribe to groups (or \textit{subreddits}) of their interests and engage with other users having similar interests.
The topics range from comedy (e.g., r/humor) to politics (e.g., r/politics) with thousands of communities of various popularity related to each topic.
The structure of the platform and its widespread influence make it ideal to study online coordinated activity.
A user can start new discussion in a community -- called a \textit{submission}.
Other users can see and reply to the submission.
Users can also reply to comments making it a thread-like structure of discussion.

For our analysis, we use data from a Russian campaign released by Reddit~\cite{reddit2017transparency} that targeted the 2016 US Presidential Elections during 2015 and 2018.
The campaign also interfered with the 2018 US Midterm Elections and 2016 Referendum.
The dataset contains 944 accounts identified by Reddit, of which 335 actively posted content. 
To obtain a dataset of accounts in the same network as these accounts, we identify accounts that reply to the troll accounts and filter their data from pre-crawled Pushshift public archive~\cite{baumgartner2020pushshift} which contains all public Reddit data from 2015 to 2020.
The complete Pushshift archive contains 600M posts and 5B comments from 2.8M subreddits.

\subsection{Twitter (X)}
Twitter (or X) is one of the most widely used online platforms for discussion on various topics ranging from politics and news to cryptocurrency.
The platform has a large audience from famous politicians and celebrities to researchers and everyday users.
The structure and widespread dissemination of content on X makes it ideal grounds to host platform manipulation efforts. 
We specifically look at the data from X's Moderation Research Consortium~\cite{twittercamps}, which contains information on platform manipulation campaigns from various state actors through the years 2018 to 2022.
The dataset is aimed to enable researchers, journalists, and the public to analyze and understand X's activities, policies, and content moderation efforts.
We pick data from 5 countries (i.e., China, UAE, Russia, Iran and Venezuela) based on popularity, activity level and documentation in past work~\cite{wang2023state}.

\descr{Ethics.} 
Our research is not considered human subjects research by our IRB since we only use publicly available data.
We follow common ethical standards and do not attempt to de-anonymize users.
We also remove personal information and usernames when presenting case studies in this work.

\begin{figure*}[!t]
	\centering
	\includegraphics[width=0.925\textwidth]{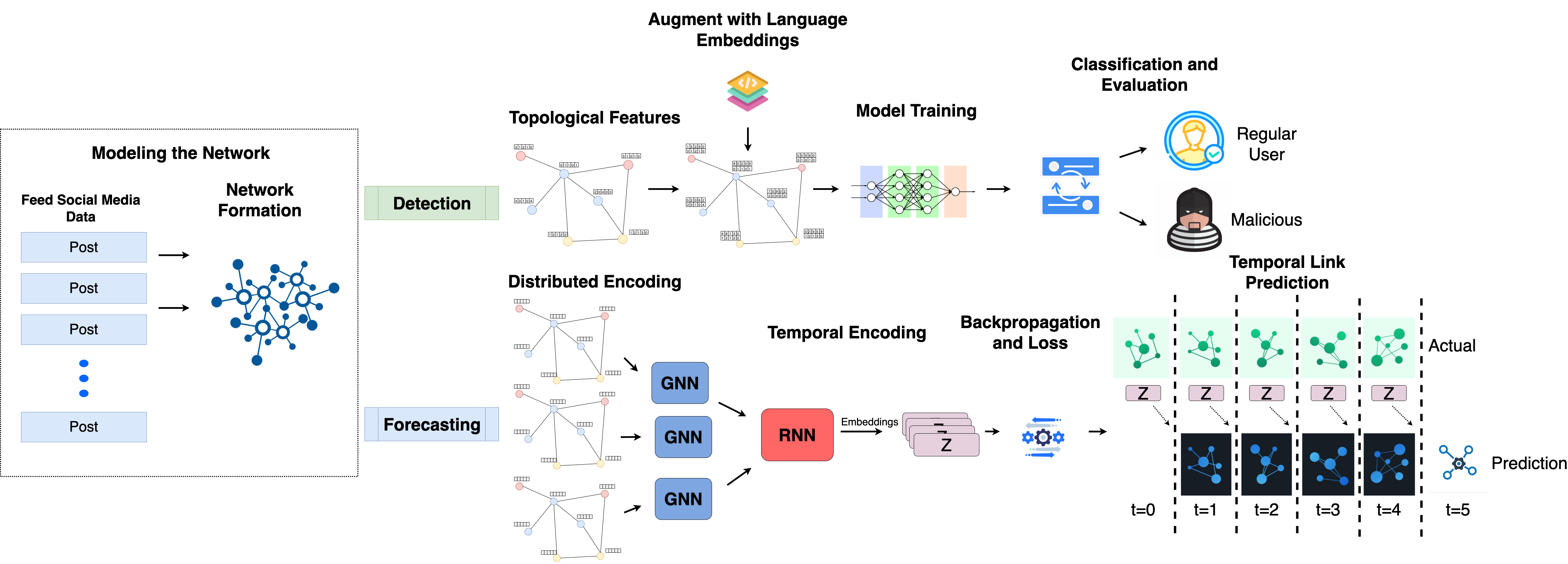}
	\caption{Overview of \name. \name first takes social media posts as input and models the network based on replying and resharing behavior. Next, it has two modes of operation: \emph{detection} and \emph{forecasting}. For detection, \name computes topological features and augments the graph with language embeddings. Then a Sage model is trained and the classifier can distinguish between regular and troll users. For forecasting, \name uses a distributed structure to scale the GNN phase and then the computed embeddings are passed through an RNN. Lastly, \name performs temporal link prediction on graph snapshots after optimization through backpropagation.}
	\label{fig:systempipeline}
\end{figure*}

\begin{figure*}[t]
\centering
\subfigure[X]{\includegraphics[width=0.45\textwidth]{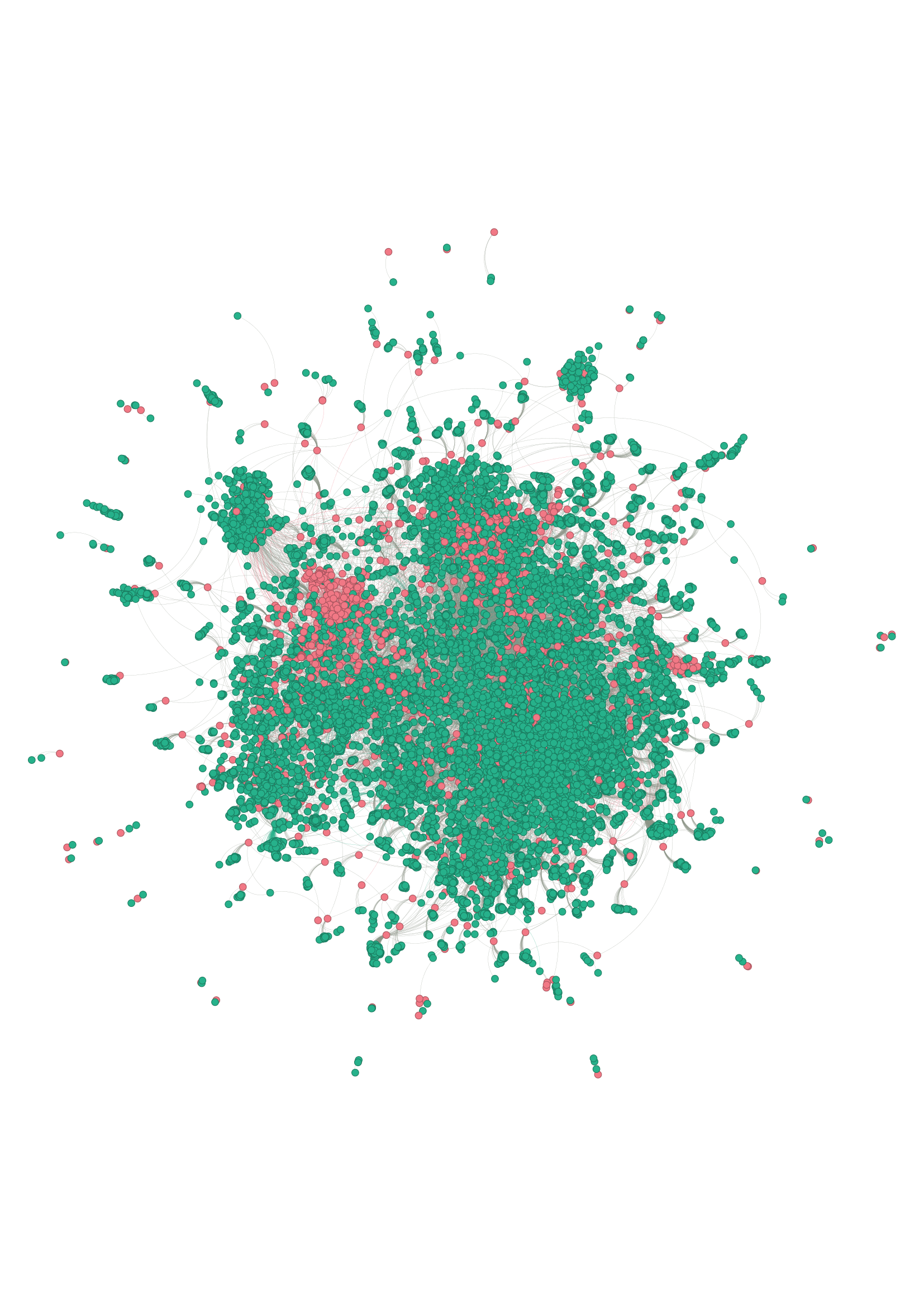}\label{fig:twitter_network}}
\subfigure[Reddit]{\includegraphics[width=0.45\textwidth]{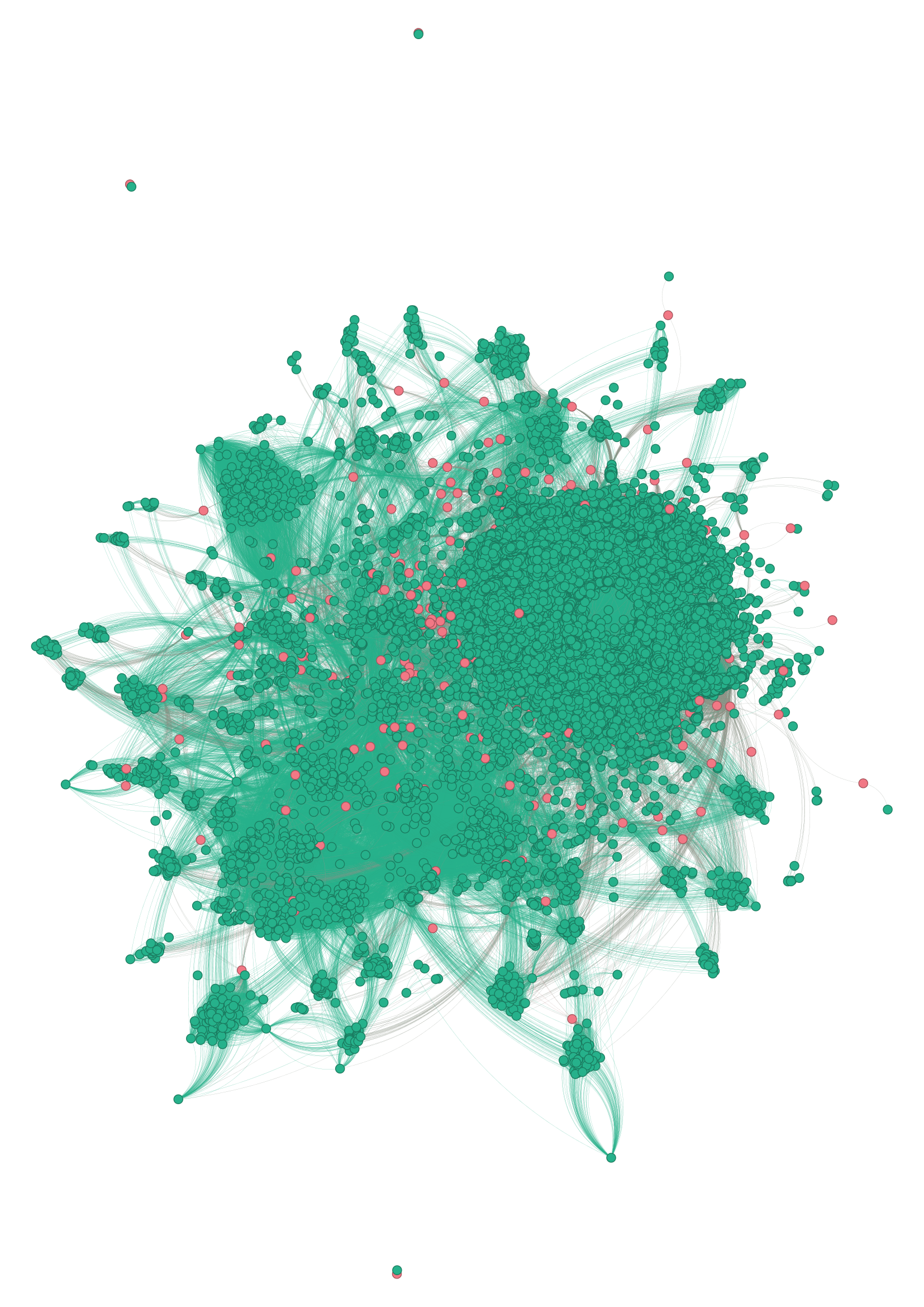}\label{fig:reddit_network}}
  \caption{Network graphs for X and Reddit. Accounts in \textcolor{red}{red} are malicious accounts and \textcolor{DarkGreen}{green} are regular users. On the left, is a UAE-based campaign on X and on the right is a Reddit campaign. }
\label{fig:network_data}
\end{figure*}

\section{Methodology}\label{sec:method}
\name uses a graph-based approach to perform two tasks: a) given a network of nodes, it identifies anomalous coordinated accounts and b) it uses temporal data to predict future links.
Figure~\ref{fig:systempipeline} highlights the complete pipeline of \name starting from modeling the network to detecting troll accounts and predicting their interactions.
In this section, we break down each stage of the pipeline.

\subsection{Data Collection}
The first step for \name is data collection.
For Reddit, the Pushshift data is structured as JSON objects with attached metadata.
Reddit has a tree-like structure, with the main post or submission having user comments and each comment can have further comments. 
To collect network data, we identify users who have replied to posts by troll accounts or reshared posts with the same title and include their data.
For X, the transparency reports provide datasets of malicious accounts as tweet objects with users they have replied to, mentioned in the tweet or retweeted.
We use that information to model the network.

\subsection{Modeling the Network}
We model the networks based on user interaction behaviors.
We define a graph \( \G = \{\V, \E\} \), where \( \V \) is the set of nodes and \( \E \) is the set of edges.  
The graph \( \G \) contains \( x = |\V| \) nodes, representing users, and \( y = |\E| \) edges, representing interactions between users.  
An edge \( (u, v) \in \E \) exists if and only if there is an interaction between user \( u \) and user \( v \), where \( u, v \in \V \).  
Thus, the graph \( \G \) models the interactions between users in the network, capturing their relationships through the edges.

For Reddit, an edge \( (u, v) \in \E \) occurs between two users if: (1) user \( v \) has replied to user \( u \)'s post or (2) user \( v \) has posted a submission with the same title (or \emph{reshared}) user \( u \)'s post after user \( u \).
For X, an edge \( (u, v) \in \E \) occurs between two users if: (1) user \( v \) replies to user \( u \), (2) user \( v \) retweets user \( u \), or (3) user \( v \) mentions user \( u \) in their tweet.

We pose the detection as an infiltration problem, where malicious actors infiltrate spaces to spread their agendas.
Therefore, when building the network graph, we consider accounts that have been reached by troll accounts (e.g., through interactions such as replies or mentions).
Using this approach, the graph has users that are close in the network space to malicious accounts.
These accounts are also the most likely to be impacted by the influence operation, allowing us to focus on communities of interest rather than processing entire Reddit data.

We use data from mentions when performing forecasting with \name.
Figure~\ref{fig:network_data} shows example networks from X and Reddit.
On the left is a UAE-based campaign from X and on the right is the Russian campaign from Reddit.
Users in red represent malicious actors and green are benign users.
The examples capture a real-world scenario, where there is a higher number of benign users within the network and a smaller number of troll accounts that have infiltrated the network in various capacities.

\descr{Generating Temporal Graphs.}
To forecast links, we model the graphs as discrete temporal graphs \( \G = \{ \G_1, \G_2, \dots, \G_T \} \).  
Each graph instance \( \G_t = \{ \V, \E_t, \mathbf{X}_t \} \) is a snapshot that captures the network state at time \( t \), where \( \V \) denotes the set of nodes in the network, \( \E_t \) is the relationships between nodes at time \( t \), and \( \mathbf{X}_t \) are the features associated with the nodes at time \( t \).
We utilize the timestamps from Reddit and X datasets to form these network graphs.

\subsection{Identifying Anomalous Coordinated Activity}
After modeling the network graph, \name involves several stages to train a GraphSage model that can identify troll accounts from a network.
We envision the detection task as node classification within a graph, where each node is a user.
Traditional machine learning approaches (e.g., supervised classification or clustering) rely on features extracted from individual entities, where the model trains to identify suspicious accounts based on activity patterns, such as the number of posts per day or the use of specific keywords. 
An advantage of Graph Neural Networks (GNNs) over traditional methods is that graph algorithms account for the relational and networked nature of influence campaigns.

\descr{Generating Features.}
\name computes the following topological features for each node in the network: \emph{in-degree}, \emph{out-degree}, \emph{degree centrality}, \emph{avg\_neighbor\_degree}, \emph{num-neighbors} and \emph{ego-net-size}.
The topological features capture the node and neighbor information to effectively perform graph model training in the next stages.
We also experiment with augmenting language embeddings from each user in the network.
The hybrid approach is meant to capture both the semantic and coordinated nature of a troll account for detection.

\descr{Model Training.} 
In order to train our system, we experiment with two popular graph algorithms, GraphSAGE and PageRank~\cite{brin1998anatomy}.
GraphSAGE computes low-dimensional vector embeddings of nodes in large graphs.
Most embedding frameworks are inherently transductive and can only generate embeddings for a single fixed graph.
In contrast, GraphSAGE is an inductive framework that leverages node attribute information to efficiently generate representations on previously unseen data.
To run GraphSAGE, it needs to train on an example graph or set of graphs.
After training, GraphSAGE generates node embeddings for previously unseen nodes or entirely new input graphs, as long as these graphs have the same attribute schema as the training data.
Similarly, PageRank is traditionally used to rank web pages, however it can also be effectively adapted for detecting malicious accounts in network-based datasets.
The algorithm assigns a ``reputation'' or ``influence'' score to each node based on the number and quality of incoming edges (links).
Since troll accounts mimic group behavior, PageRank can detect unusual patterns in the graph.

\subsection{Temporal Link Prediction}
\name adopts a scalable temporal link prediction strategy discussed in~\cite{king2023euler}.
We stack a GNN upon a RNN sequence encoder.
In this approach, the network topology at time \( t\) is encoded by the GNN and the dynamic connections are modeled by the RNN.
\name replicates the GNNs across several worker machines so that snapshots can be computed concurrently and independently.
This method allows a significant speed-up and offers scalability for millions of nodes and edges making it ideal for real-world deployment.

\section{Results}

\subsection{Detection of accounts (RQ1)}
Our best implementation for Reddit uses GraphSAGE and language embeddings generated through OpenAI's TE3-S (or \textit{text-embedding-3-small}) model  followed by Sentence BERT (or SBERT)~\cite{reimers2019sentence}.
The \textit{text-embedding-3-small} model belongs to the family of OpenAI's embedding models and achieves better performance on benchmark tasks than the earlier \textit{text-embedding-ada-002} model.
We use the small version keeping in view the compute, memory and storage cost.
Additionally, SBERT is also particularly well-suited for this task since it generates sentence-level embeddings and offer multi-lingual support.
The embeddings are generated per user and are coded as node features along with the degree features.
As seen in Table~\ref{tbl:reddit_results}, our results show that using SAGE without language embeddings still outperform the Interaction based features which are state-of-the-art for Reddit dataset~\cite{saeed2022troll}.
The complete graph for Reddit includes 46k nodes and 273k edges.
As discussed in Section~\ref{sec:method}, the graph contains nodes of interest that have interacted with troll accounts.
This approach enables us to distinguish troll accounts from users that are close in the network space, are likely to be the most impacted and contribute to similar topics.
The interaction based features for past work include: \emph{Total Comments}, \emph{Total Submissions}, \emph{Account Age}, \emph{Fraction of submissions with the same title as troll accounts}, \emph{Fraction of comments on submissions that troll accounts
commented}, \emph{Fraction of comments on submissions by troll accounts}, \emph{Fraction of direct replies on submissions by troll accounts}, \emph{Fraction of comments that are a reply to a troll account's comment}, and \emph{Fraction of comments that are a reply to a troll account's comment in a troll account's submission}.
We use the output embeddings from the graph neural network for downstream classification through a Random Forest classifier with a 80-20 training/test split.
To ensure robustness of our results, we perform 10-fold cross-validation similar to~\cite{saeed2022troll} to ensure a fair comparison with the baseline.

We replicate our findings on 5 popular campaigns on X and compare our approach with PageRank and a standard user-information based classifier.
Due to a lack of tweet data for benign users, we do not incorporate language embeddings and rely solely on topological features.

\begin{table*}[t!]
\centering
\caption{Performance of different detection approaches on Reddit network.}
\begin{tabular}{lcccc}
\toprule
\textbf{Method} & \textbf{AUC} & \textbf{Precision} & \textbf{Recall} & \textbf{F1} \\
\midrule
Interaction & 90.29 $\pm$ 1.90 \% & 95.33 $\pm$ 1.65 \% & 90.29 $\pm$ 1.90\% & 92.74 $\pm$ 1.27\% \\
SBERT & 50.69 $\pm$ 0.68\% & 69.38 $\pm$ 17.77\% & 50.69 $\pm$ 0.68\% & 58.58 $\pm$ 6.35\% \\
PageRank & 80.47 $\pm$ 0.86\% & 50.96 $\pm$ 0.02\% & 80.47 $\pm$ 0.86\% & 62.40 $\pm$ 0.26\% \\
SAGE & 92.17 $\pm$ 2.06\% & 96.71 $\pm$ 1.53\% & 92.17 $\pm$ 2.06\% & 94.38 $\pm$ 1.30\% \\
SAGE + SBERT & 93.61 $\pm$ 2.02\% & 97.97 $\pm$ 1.30\% & 93.61 $\pm$2.02\% & 95.74 $\pm$ 1.23\% \\ 
SAGE + TE3-S & \textbf{93.70} \bm{$\pm$} \textbf{1.91\%} & \textbf{99.36} \bm{$\pm$} \textbf{0.79\%} & \textbf{93.70} \bm{$\pm$} \textbf{1.91\%} & \textbf{96.44} \bm{$\pm$} \textbf{1.08\%} \\
\bottomrule
\end{tabular}
\label{tbl:reddit_results}
\end{table*}

\descr{Implementation Details.}
\name's implementation has a 2-layer GNN and uses ReLU activation function.
The pipeline is built on PyTorch for GNNs and NetworkX to model the graph.
We optimize \name by varying the number of epochs [10, 100, 1000] and learning rates [0.1, 0.01 and 0.001].
We report all the results for the best configuration of \name that runs for 100 epochs with a learning rate of 0.001 using Adam optimizer~\cite{pytorch_ddp} and CrossEntropy as the Loss function.
\name uses a pre-trained SBERT (paraphrase-multilingual-MiniLM-L12-v2 model) which takes as input a user's posts and outputs a 384-size vector embedding per user.
Similarly, the TE3-S model in Table~\ref{tbl:reddit_results} refers to OpenAI's text-based embedding model, more specifically, the \textit{text-embedding-3-small} model.
Messages are processed in batches to respect API constraints, with embeddings computed for each message.
The output dimensions for this model are 1536.
We feed the most recent 100 posts per user to generate the embeddings.
To respect the token limit of the model, we feed the posts in chunks and average the results to obtain the final embeddings.

To train the standard baseline user-information classifier, we use 500 benign accounts randomly selected from a 2022 pre-crawled Twitter 1\% data stream and the (now) obsolete Twitter Search API for their data.
Popular user-information features (from past research~\cite{mihaylov2019hunting, varol2017online, echeverri2018lobo}) include: \emph{Tweet Count}, \emph{Account Age}, \emph{Followers}, \emph{Following}, \emph{Account Language}, \emph{Account Description Language} and \emph{Description Length}, \emph{Average mentions per Tweet}, \emph{Average Tweet Length} and \emph{Fraction of Tweets that are Retweets}.
As seen in Table~\ref{tbl:graph_classification}, SAGE outperforms the other two methods, namely PageRank (PR) and Random Forest (RF), for most metrics and is well suited for this task.
With the exception of higher precision in PageRank for certain campaigns (e.g., Iran and Russia).
Since PageRank classifies based on ``influence'' or measure of importance within the network, it is likely that troll accounts from these campaigns were particularly influential. 
A possible strategy from this finding is to augment the SAGE based approach with additional PageRank scores, however there is a high chance of overfitting since both are graph-based approaches.

\begin{figure*}[t]
    \centering
    \subfigure[Robustness testing by adding noise to various networks.\label{fig:noise}]{
        \includegraphics[width=0.46\textwidth]{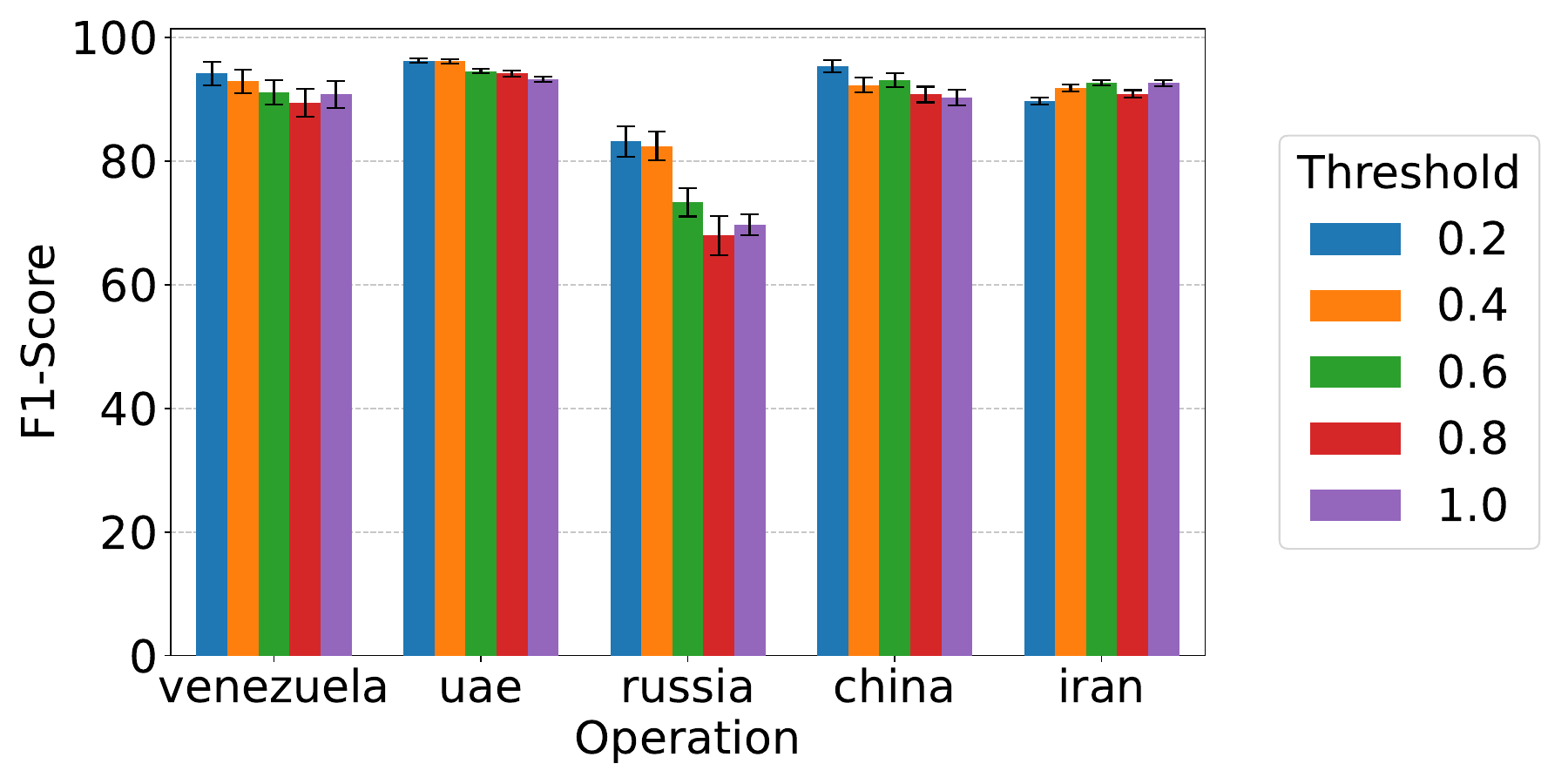}
    }
    \subfigure[Ablation testing on various networks.\label{fig:ablation}]{
        \includegraphics[width=0.48\textwidth]{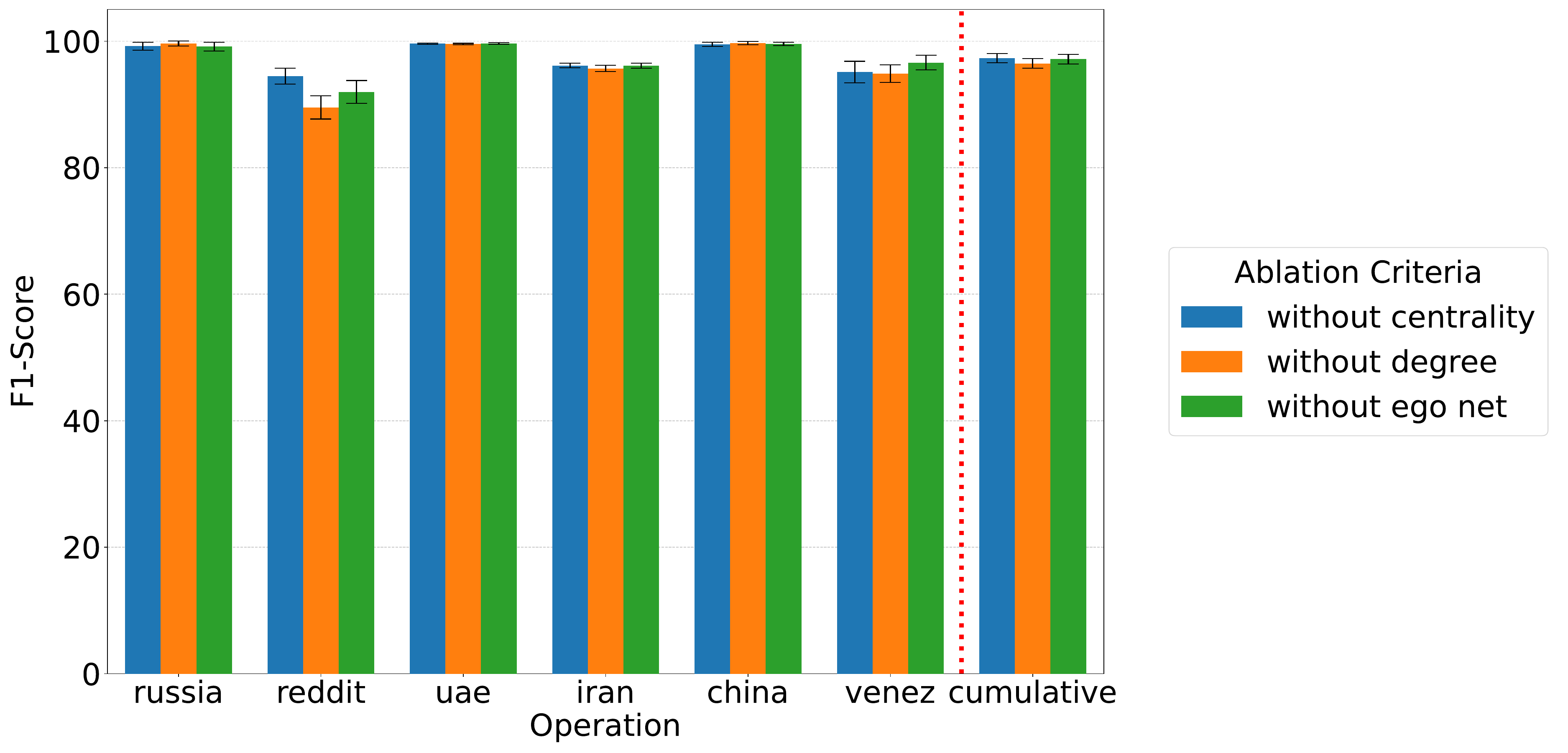}
    }
    \caption{Results from ablation and robustness testing on X network.}
\end{figure*}

\descr{Robustness Testing.}
We augment the edges in the graphs using the Barabasi-Albert (BA) algorithm~\cite{barabasi2002stat} to the regular users in an attempt to dilute the troll interactions for robustness testing.
The algorithm has been extensively used in past research for modeling social media graphs~\cite{burbach2019shares, rabb2022cognitive, rajabi2020modeling}.
Our results are highlighted in Figure~\ref{fig:noise}.
As it stands, with increasing number of benign edges, the model is still able to capture nuances and intricate coordinated patterns within the malicious groups.

\begin{table*}[t!]
    \begin{center}
        \caption{Performance of different detection approaches on X network.}
            \label{tbl:graph_classification}
        \begin{tabular}{lrrrrrrr} 
            \toprule
            {\bf Campaign} & \small {\bf Activity} & \small {\bf Graph} &  \small {\bf Method} & \small {\bf AUC} & \small {\bf Precision} & \small {\bf Recall} &\small {\bf F1}\\
            \midrule
            \multirow{3}{*}{China} & 2008 to   & 48.3k nodes \&  & SAGE & \textbf{99.63 \bm{$\pm$} 0.34\%} & 99.60 $\pm$ 0.32 & \textbf{99.62 \bm{$\pm$} 0.34\%} & \textbf{99.61 \bm{$\pm$} 0.27\%} \\ 
               &  2019  &         55.7k edges                & PR & 93.15 $\pm$ 0.66\% & \textbf{99.93 \bm{$\pm$} 0.006\%} & 93.16 $\pm$ 0.66\% & 96.42 $\pm$ 0.35\% \\ 
               &     &                        & RF  & 61.23 $\pm$ 2.81 & 61.27 $\pm$ 2.81 & 61.23 $\pm$ 2.81 & 61.25 $\pm$ 1.98 \\ \hline
            \multirow{3}{*}{Venezuela} & 2015 to   & 367 nodes \&  & SAGE & \textbf{95.41 \bm{$\pm$} 2.41\%} & \textbf{95.56 \bm{$\pm$} 2.39\%} & \textbf{95.40 \bm{$\pm$} 2.41\%} & \textbf{95.48 \bm{$\pm$} 1.70\%} \\ 
                   &     2018      &     387 edges       & PR & 88.19 $\pm$ 1.06\% & 90.99 $\pm$ 0.67\% & 88.19 $\pm$ 1.06\% & 89.57 $\pm$ 0.4\% \\ 
                   &           &            & RF  & 62.20 $\pm$ 2.96\% & 62.23 $\pm$ 2.96\% & 62.20 $\pm$ 2.96\% & 62.21 $\pm$ 2.09\% \\ \hline
            \multirow{3}{*}{Iran} & 2009 to   & 180.6k nodes \&  & SAGE & \textbf{97.47 \bm{$\pm$} 0.50\%} &  94.77 $\pm$ 0.59\% & \textbf{97.47 \bm{$\pm$} 0.50\%} & \textbf{96.10 \bm{$\pm$} 0.39\%} \\ 
              &         2018       &       640.9k edges         & PR & 84.22 $\pm$ 0.38 & \textbf{99.41 \bm{$\pm$} 0.06} & 84.22 $\pm$ 0.38 & 91.19 $\pm$ 0.23 \\ 
              &                &                & RF  & 76.61 $\pm$ 2.58\% & 77.24 $\pm$ 2.72 & 76.61 $\pm$ 2.59\% & 76.92 $\pm$ 1.88 \\ \hline
            \multirow{3}{*}{Russia} & 2010 to   & 100.8k nodes \&  & SAGE & \textbf{99.10 \bm{$\pm$} 0.81\%} & 98.86 $\pm$ 1.06\% & \textbf{99.09 \bm{$\pm$} 0.81\%} & \textbf{98.98 \bm{$\pm$} 0.67\%} \\ 
                &         2018          &       138.1k edges      & PR & 89.04 $\pm$ 1.04\% & \textbf{99.97 \bm{$\pm$} 0.002\%} & 89.04 $\pm$ 1.04\% & 94.19 $\pm$ 0.58\% \\ 
                &                   &             & RF  & 60.36 $\pm$ 2.91 & 60.52 $\pm$ 2.94 & 60.36 $\pm$ 2.91 & 60.44 $\pm$ 2.07 \\ \hline
            \multirow{3}{*}{UAE}	& 2011 to   & 35.7k nodes  \&  	&	SAGE	&	\textbf{99.54 \bm{$\pm$} 0.17\%}  & \textbf{99.69 \bm{$\pm$} 0.16\%} & \textbf{99.55 \bm{$\pm$} 0.17\%} & \textbf{99.61 \bm{$\pm$} 0.16\%} \\
            &	2019 & 83.4k edges &	PR	&	84.66 $\pm$ 0.23\%  & 98.58 $\pm$ 0.02\% & 84.66 $\pm$ 0.23\% & 91.09 $\pm$ 0.13\%\\
            & & &	RF 	& 86.30 $\pm$ 1.92\%  & 87.10 $\pm$ 1.80\% & 86.30 $\pm$ 1.92\% & 86.70 $\pm$ 1.32\%\\ \hline      
            \bottomrule
        \end{tabular}
    \end{center}
\end{table*}

\descr{Ablation Testing.}
We also perform an additional ablation study on the Sage model without language embeddings to confirm the most meaningful topological features for the detection task.
Our results are highlighted in Figure~\ref{fig:ablation}.
We find that different campaigns have slightly different outcomes.
However, for Reddit specifically, we observe that degree features are the highest contributors to performance.
We also find that degree features cumulatively have the highest impact, followed by ego net and lastly centrality.

\subsection{Forecasting Malicious Links (RQ2)}
Graph neural networks perform inference best when they have informative features about the nodes in a network, especially GraphSAGE~\cite{gnn_feature_importance}. 
Unfortunately, our dataset only contains information about troll users, meaning all non-trolls would appear identical to the GNN. 
To remedy this, \name generates several topological features in an attempt to summarize each user's behavior within the network. 
We start with one-hot labels of the user's type: troll or non-troll. 
\name then computes additional topological features about each node's neighborhood. 
First, it aggregates the labels of neighboring nodes using message passing to create a feature for the sum and mean of each source node in the user's one-hop neighborhood. 
Then, it repeats this process for destination nodes, and again to generate features if the graph were undirected. 
Finally, as before, we calculate each node's degree centrality measure:

\begin{equation}
    \label{eq:degree_centrality}
    d(v_i\mid \G_t) = \frac{|\mathcal{N}(v_i)|}{\underset{v_j \in \V_t}{max} \{|\mathcal{N}(v_j)|\}}.
\end{equation}
This information summarizes the proportion of trolls and non-trolls the users tend to interact with over a given period of time, as well as the number of trolls and non-trolls that interact with them and how active they are compared to other users generally.
This allows us to build an informative profile about each user that the GNN can better learn from. 

In the next step, \name partitions the edge list into temporally even chunks, each spanning $\delta$ units of time. 
For each dataset, we selected the smallest possible value of $\delta$ such that no snapshot had fewer than 16 edges. 
Using $\delta=7$ days was sufficient for all datasets.
For forecasting, we scale the analyses and account for multiple campaigns originating from the countries.
We also include mentions, resulting in much larger graphs for each campaign.
Table~\ref{tab:euler_meta} shows the metadata of this data partition. 

\begin{table}[t]
\caption{Link Prediction Metadata}
\label{tab:euler_meta}
\begin{tabular}{lrrrr}
    \toprule
     Campaign& $|\V|$ & $|\E|$ & Trolls  & Avg. Chunk Size \\\midrule 
     China & 376k & 3.52m & 940 & 5.8k\\
     Venezuela & 382k & 6.94m & 1951 & 15k\\
     Iran & 414k & 6.17m & 2311 & 14k\\ 
     Russia & 234k & 2.04m & 416 & 5.3k\\
     UAE & 106k & 1.06m & 4248 & 2.8k \\
     Reddit & 46k & 273k & 288 & 830\\
     \bottomrule 
\end{tabular}
\end{table}

\begin{table*}[h!]
    \begin{center}
    \caption{Temporal Link Prediction Performance for troll accounts on X and Reddit.}
        \label{tbl:link_detection}
        \begin{tabular}{lrrrrrrrr} 
            \toprule
            {\bf Campaign} & \small {\bf Method} &  \small {\bf AUC} & \small {\bf AP} & \small {\bf TTE-AUC} & \small {\bf TTE-AP} & \small {\bf TUE-AUC} & \small {\bf TUE-AP}\\
            \midrule
            \multirow{ 2}{*}{China} & SAGE + RNN & \textbf{0.9692} & \textbf{0.9736} & \textbf{0.8508} & \textbf{0.9583} & \textbf{0.9694} & \textbf{0.9740}\\
            & SAGE & 0.6813 & 0.6858 & 0.6415 & 0.8413 & 0.6813 & 0.6849 \\ \hline
            \multirow{ 2}{*}{Venezuela} & SAGE + RNN & \textbf{0.9809} & \textbf{0.9817} & \textbf{0.9449} & \textbf{0.9859} & \textbf{0.9823} & \textbf{0.9833} \\ 
            & SAGE & 0.5298 & 0.5989 & 0.4715 & 0.8248 & 0.5295 & 0.5957 \\ \hline
            \multirow{ 2}{*}{Iran} & SAGE + RNN & \textbf{0.9643} & \textbf{0.9759} & \textbf{0.9278} & \textbf{0.9974} & \textbf{0.9645} &  \textbf{0.9760}\\
            & SAGE & 0.5996 & 0.7726 & 0.6307 & 0.9864 & 0.5990 & 0.7688 \\ \hline
            \multirow{ 2}{*}{Russia} & SAGE + RNN & \textbf{0.9889} & \textbf{0.9867} & \textbf{0.9369} &  \textbf{0.9590} & \textbf{0.9894} & \textbf{0.9874}\\
            & SAGE & 0.5036 & 0.5987 & 0.5022 & 0.5742 & 0.5037 & 0.5987 \\ \hline
            \multirow{ 2}{*}{UAE} & SAGE + RNN & 0.8923 & 0.9010 & \textbf{0.9194} & \textbf{0.9552} & 0.8878 & 0.8955 \\
            & SAGE & \textbf{0.9110} & \textbf{0.9117} & 0.9142 & 0.9551 & \textbf{0.9078} & \textbf{0.9042} \\ 
            \midrule 
            \multirow{2}{*}{Reddit} & SAGE + RNN & 0.9997 & 0.9978 & \multicolumn{1}{c}{\hspace{1em}--} & \multicolumn{1}{c}{--} & 0.9997 & 0.9978 \\
            & SAGE & \textbf{0.9999} & \textbf{0.9999} & \multicolumn{1}{c}{\hspace{1em}--} & \multicolumn{1}{c}{--} & \textbf{0.9999} & \textbf{0.9999} \\
            \hline
            \bottomrule
        \end{tabular}
    \end{center}
\end{table*}

The models were optimized to predict links in $\G_{t+1}$, given the node embeddings of $\G_t$. 
First, we calculate the features of $\G_t$. 
Next, the model, either a GNN or a GNN and an RNN process the features and edges in the graph to produce node embeddings, $\mathbf{Z}_t$. 
The likelihood of an edge $(u,v)$ in $\G_{t+1}$ is then calculated as 

\begin{equation}
    \label{eq:link_pred}
    P((u,v) \in \G_{t+1} \mid \mathbf{Z}_t) = \sigma \Big( \mathbf{Z}_t[u] \cdot \mathbf{Z}_t[v] \Big)
\end{equation}
where $\sigma(\cdot)$ is the sigmoid function. 

\descr{Implementation Details.}
\name uses 2-layer GNNs with 64-dimensional hidden layers. 
The models with RNNs also used 64 hidden dimensions, and the models without RNNs used an additional 64-dimensional linear layer to ensure the same number of parameters between experiments. 
Finally, the output is projected to 32-dimensional embedding vectors which are used for link prediction. 

In each experiment, we distribute the work of the GNNs across 16 workers, and a single coordinating process. 
When models use RNNs, the output of the workers is passed into the RNN hosted on the coordinator. 
Finally, the coordinator sends the embeddings for $\mathbf{Z}_{t-1}$ and the edges for $\G_t$ to each worker to calculate loss. 
This distributed approach allows our models to scale to the million-edge datasets they need to process, and train in a matter of hours on only CPUs. 
We utilize a maximum of 64 CPUs and our setup is an Intel(R) Xeon(R) Gold 6338 CPU @ 2.00GHz with 251GB RAM. 

In all experiments, we partitioned the data in temporal order. 
Models were trained on the first 80\% of the snapshots, validated using the next 5\%, and tested on the final 10\%. 
To evaluate the models, we use the set of unique edges in the next snapshot as the set of positive samples, and use the same number of random edges as the negative sample. 
The models train until average precision stops improving on the validation set for 10 epochs. 
In all experiments, we used a learning rate of 0.001, and the distributed Adam optimizer implemented in PyTorch~\cite{pytorch_ddp}.
\name predicts all links from the test period, allowing for the detection of both troll–troll and troll–user edges.

\descr{Results.}
Table~\ref{tbl:link_detection} shows the results of these experiments. 
We observe that overall, the SAGE+RNN model performs very well in detecting future interactions between trolls and users on X and Reddit. 
In all experiments, for both metrics it scored in the high 90s. 
The SAGE+RNN model scored significantly higher than the model without the RNN in all datasets except the UAE campaign and the Reddit dataset.

Notably, during the Russian and Venezuelan campaigns, the difference between the RNN models that take time into account and those that do not is most evident. 
The model that only used GraphSAGE scored almost as poorly as random (AUC $\approx 0.5$) while the model that included an RNN remains highly accurate. 
It may be because these campaigns have more edges per unit time than the campaigns this approach scored well on. 
As a result, this would make the changes in user behavior over time more evident, or due to the larger size of the snapshots, make solely structural analysis insufficient for prediction. 
This idea is supported by how well the SAGE model, even without the support of RNN, performs on the UAE and Reddit datasets, which had fewer edges overall. 
On the Reddit dataset, the SAGE model actually outperforms the model that includes an RNN, though both score $>0.99$ in both metrics. 

To answer \textbf{RQ2} with more granularity, we evaluate the link predictors on troll-to-troll edges (TTE), and troll-to-user edges (TUE) separately. 
Models that excel at predicting TTEs would allow researchers to potentially forecast the coordination of these campaigns; models that accurately predict TUEs could allow social media companies to warn susceptible users that they are at risk of being taken advantage of. 
With the exception of the UAE campaign, the models are best at predicting links between trolls and non-trolls. 
The outlier of this campaign is likely explained by the fact that it had so many more troll users than the others, both proportionally, and by count. 
When trolls are more decentralized, as is the case in the test partition of the Reddit dataset, there are no TTEs present, so the ability to predict TUEs is especially important. 
Additionally, models must be good at predicting these kinds of links because they are the more insidious types of interactions that can occur. 
In an upcoming section, we will highlight some interactions Trolls had with normal users that our models were able to forecast to further motivate why the ability to predict these edges is so important.
The high performance on the Reddit dataset can be attributed, in part, to the dataset's dominance by a single, well-studied troll campaign (i.e., the Russian influence operation), reducing task complexity.
Additionally, the behavioral separation between trolls and regular users in this dataset (e.g., coordination patterns~\cite{saeed2022troll}) allows the model to rely on more discriminative signals.

\section{Case Studies}
When predicting interactions, \name identifies the edges of spreaders with other spreaders (or \textit{troll-to-troll edges}) and also the interaction of spreaders with regular users (or \textit{troll-to-user edges}).
In essence, troll-to-user edges refer to \textit{targeted influence} where malicious users tailor content to manipulate specific groups and troll-to-troll edges mimic \textit{amplification} of content where they inflate and reshare each others posts or comment underneath to increase visibility.


Both of these strategies have been studied discussed in prior works.
For amplification, prior research highlights tactics such as repetitive posting and emotional appeals~\cite{shao2018spread}, use of hyperbolic or sensational claims~\cite{zannettou2019web}, and coordinated reply campaigns designed to drown out opposing content~\cite{starbird2019disinformation}.
Similarly, for target influence, some works have shown Russian trolls posing as Black activists (e.g., \@Blacktivists) to incite racial tensions~\cite{badawy2018analyzing}, curating content to align with audience interests~\cite{boatwright2018troll}, and using culturally specific language such as local slang~\cite{cinelli2022coordinated}.
We aim to highlight instances of both amplification and targeted influence through the following case studies.
Each example is an edge (or interaction) predicted by \name, demonstrating the potential to curb such tactics before they reach broader audiences.


\descr{Amplification.}
One of the most effective ways to spread a narrative is through boosting, which is why resharing patterns are the cornerstone of the most detection systems built for online campaigns.
Below are some examples of tweets that were being inflated (through reshares) by troll accounts and were preemptively detected by \name.

\begin{mdframed}[style=lightgrayframe]
\textbf{UAE Campaign}

\noindent \textbf{Example:} \#Cohen denies `direct evidence' of \#Russia collusion; accuses [REDACTED] to be involved in Roger Stone- \#WikiLeaks plot 

\noindent \textbf{Russian Campaign}

\noindent \textbf{Example:} WATCH: Hillary Said In 2010 That Obama Wanted To `Strengthen Russia' - [REDACTED LINK] \#DeepStateCorruption \#Democ
\end{mdframed}

The examples show certain political statements being inflated which is consistent with actions of troll accounts in past work (such as excessively retweeting narratives that back their agendas~\cite{stewart2018examining}).

\descr{Targeted Influence.}
We find that targeted influence is more \emph{real}, such that troll accounts communicate like any other user to do so.
Hence, we argue that while amplification can be as simple as retweeting news, targeted influence is more nuanced and even more difficult to ascertain.
Here are a few examples of targeted influence captured by \name:

\begin{mdframed}[style=lightgrayframe]
\noindent \textbf{Russian Campaign} 

\noindent \textbf{Example:} This country is way behind on intellectuals and wise ppl., otherwise we wouldn't have [REDACTED] as president. -\#facts

\noindent \textbf{Example:} @[REDACTED USER] Undoubtedly, @[REDACTED NAME] is the most desired-to-be-seen-playing football in America. After the NFL's long black-balling, I think the world can't wait to see this quarterback get back onto the playing field.
Of course except for these weak-ass white supremacists.

\noindent \textbf{UAE Campaign} 

\noindent \textbf{Example:}  @[REDACTED USER] someone here believes that [REDACTED] make for African Americans more than the previous black president!

\noindent \textbf{Iranian Campaign} 

\noindent \textbf{Example:}  @[REDACTED USER]: It's a complete lie that \#President's son did not know that this was a Russian government effort to provide information about Clinton because the email itself says in black;  white this is part of Russian government's support for [REDACTED]; his campaign. \#AMJoy
\end{mdframed}
The tweets highlighted above are directed at specific users on the platform and were identified as troll-user edges by \name.
Malicious actors often craft identities and engage directly with real users~\cite{linvill2020troll}, specifically in groups of 2 or 3~\cite{saeed2022troll}, making their influence more persuasive and potentially more harmful.
The examples show how troll accounts personify to convince users of their agendas making it crucial for detection and forecasting systems to operate at a fine-grained level.

\subsection{Remarks}
Due to the nature of this research, the presented case studies can be offensive or condescending to the readers.
The aim is to highlight the shades of content and the granularity with which influence happens in certain spaces.
Ideally, platforms should represent organic conversation between users with appropriate moderation where necessary.
As the scale of platforms increase, so does the opportunity for manipulation, which makes it important to develop systems that can curb malicious activity efficiently and in a timely manner.

\section{Related Works}
\descr{Graph based approaches.}
Graph-based approaches have been successfully used for a variety of tasks in social media, ranging from fake news detection to node classification.
For example, using GNNs substantially improves fake news detection performance as compared to purely text-based models~\cite{chandra2020graph, wang2020fake, mohawesh2023semantic}.
In~\cite{pierri2020multi}, the authors classify credible vs non-credible news with an AUROC up to 94\% by using diffusion networks of retweets and mentions using network features (e.g., number of connected components) to perform classification. 

Disinformation detection is also tackled using various graph approaches, e.g., hypergraphs to perform binary classification of retweet cascades into fake and not fake classes~\cite{salamanos2024hypergraphdis} and using network structure features in addition to post content and user sentiment on X~\cite{paraschiv2022unified}.
For node classification tasks,~\cite{pho2022exploring} shows Simple Graph Convolution (SGC) applied to the Facebook Page-Page dataset.
In~\cite{michail2022detection}, the authors create a graph based model that predicts whether the original tweet and its diffusion thread constitute an organized astroturfing campaign, where the campaigns are loosely defined as retweet trees. 

Hamid et al.~\cite{hamid2020fake} use Graph Neural Networks (GNNs) for structure-based fake news detection and achieve an average ROC of .95\%.
Huang et al.~\cite{huang2019deep} present a hybrid neural network model for rumor detection on X, which is the first work that models user with graph convolutional networks for rumor detecting and achieves much better performance than the state-of-the-art methods.
Song et al.~\cite{song2021temporally} introduce a novel temporal propagation-based fake news detection framework, which could fuse structure, content semantics, and temporal information to model temporal evolution patterns of real-world news.

\descr{Troll Account Detection.} 
There are several works that focus on detection of troll accounts on social media platforms.
For example, Alhazbi et al.~\cite{alzahbi200mach} use behavioral features to detect state-sponsored trolls on X.
They use a dataset of Saudi trolls disclosed by X in 2019 to train and evaluate their system, achieving an accuracy of 94.4\%.
Similarly, Fornacciari et al.~\cite{forn2018holistic} proposes TrollPacifier, a system that uses six groups of features for detection: writing style, sentiment, behaviors, social interactions, linked media, and publication time.
Their system detects state-sponsored trolls on X using a neural network model with an accuracy of 95.5\%.
To capture the coordinated nature of troll accounts, we use a graph-based approach for detection tasks and compare \name to various baselines in our experiments.

\descr{Influence Campaigns.}
Influence operations and platform manipulation efforts have been widely studied in the literature~\cite{bessi2016social,ferrara2017disinformation, ferrara2016rise,varol2017online}.
This research uncovers the role of social bots in the spread of information regarding political events and the extent to which they manipulate public opinion during significant events (e.g., elections).
Other works, such as Mihaylov et al.~\cite{mihaylov2016hunting} dig deeper and highlight the different kinds of accounts involved for influence.
They find that accounts responsbile for influence operations are either independent actors or financially incentivized actors that perform targeted influence.
Similarly, other works such as Stewart et al.~\cite{stewart2018examining} highlight the motives behind campaigns and show that Russian-sponsored trolls engage in Black Lives Matter (BLM) discussion on X in an attempt to promote narratives targeted towards both left and right-leaning communities, while Zannettou et al.~\cite{zannettou2020characterizing, zannettou2019disinformation,zannettou2019let} conduct a variety of studies examining state-sponsored trolls operating on X and Reddit between 2014 and 2018 and evaluate their effectiveness in disseminating content on the platforms.
In an attempt to curb influence operations, Ratkiewicz et al.~\cite{ratkiewicz2011detecting} use machine learning techniques to identify the spread of false political information on X.

Existing graph methods to detect malicious accounts focus mainly on fake accounts and social bots.
Li et al.~\cite{li2022sybilflyover} present SybilFlyover, which is a heterogeneous graph-based model to detect fake accounts.
The authors test it on the My Information Bubble (MIB) dataset~\cite{cresci2015fame} of fake Twitter followers.
Similarly, Liu et al.~\cite{liu2024segcn} develop a subgraph encoding based graph model which uses convolutional neural networks to detect social bots on Twitter.
In the link prediction domain, Xue et al.~\cite{xue2021modeling} present DyHATR, which is a heterogeneous network embedding method evaluated on social media datasets for link prediction.
However, past work has shown that troll accounts behave differently from social bots, spammers and fake accounts~\cite{mueller2019mueller, zannettou2019let, zannettou2019disinformation, saeed2022troll}.
These behaviors include, but are not limited to, coordinated posting, adopting opposing sides of debates, and mimicking regular user activity.
Therefore, systems designed to detect fake or malicious accounts are often ineffective against troll accounts.

For detecting troll accounts using GNNs, Minici et al.~\cite{minici2025iohunter} present IOHunter that detects accounts from influence campaigns on X.
The authors train a multi-modal fusion model utilizing text embeddings in a graph foundation model.
Similarly, Asif et al.~\cite{asif2024graph} utilize GloVe word embeddings in combination with Graph Convolutional Networks (GCNs) to detect troll behavior and content on X. 
In contrast to these approaches, our detection module treats text embeddings as node features and decouples graph embedding learning from classification.
This design enables generalization and allows us to isolate and measure the impact of the graph-based approach more clearly on two different platforms (i.e., Reddit and X).
Additionally, \name leverages a distributed framework that decomposes graphs into temporal snapshots and forecasts future troll–troll and troll–user interactions.

\section{Discussion and Conclusion}
We present \name, a system that utilizes social media network data to identify anomalous troll accounts and predict their future actions.
This work paves the way for improved detection of troll accounts and  identification of malicious influence over users on social media.
Our goal is to increase user safety and improve trust in the online social media ecosystem.
We adopt a multi-platform approach and perform analysis on datasets provided by both Reddit and X.
We find that \name's multi-modal pipeline using Sage with topological features and language embeddings outperforms state-of-the-art node detection scores for Reddit campaign offering a 3.7\% improvement in F1-score and generalizes well with high performance over X operations.
We also show via \name's temporal forecasting pipeline utilizing a combination of GNNs and RNN that we can predict future \emph{troll-troll} and \emph{troll-user} interactions with an average AUC of 96.6\%.
Through our work, we aim to highlight an effective combat strategy against modern influence campaigns.

\subsection{Broader Impacts}
For a fight against malicious influence, social media research has explored various facets from detection to measurement efforts.
Modern influence campaigns are spread out on various platforms and originate from different countries.
As platform manipulation grows in scale, preemptive detection and forecasting of malicious behavior are increasingly important, since online platforms can present harm for the end user in various ways, e.g., being led into conspiratorial echo chambers, or being targeted online for being of specific racial backgrounds, to name a few.


Since social media datasets are large in size, \name relies on GraphSAGE which enables scalable learning on large graphs and is suitable for evolving graphs.
For link prediction, \name implements a distributed training architecture with 16 workers and a single coordinating process.
Each worker computes local GNN embeddings, while the coordinator hosts any RNN components (if temporal modeling is used) and handles the aggregation of embeddings and edge information.
This allows us to efficiently process million-edge graphs using only CPUs, making it practical for deployment in real-world, large-scale applications.

\descr{Responsible Deployment.} Ultimately, \name is a tool for mitigating harmful online behavior. 
However, we acknowledge the associated implications, such as false positives and erroneous classification.
In this context, false positives are a known challenge, particularly in moderation systems where errors can stifle free expression.
While our approach makes best efforts to minimize them, there is no perfect solution.
We encourage deployment in a manner that combines \name with human judgment, ensuring that decisions are made responsibly.
We also envision defining confidence thresholds and only escalating highly confident decisions, where lower-confidence cases can be monitored without direct action and users flagged by the system should have accessible avenues to contest decisions, with clear explanations of triggering behaviors.

However, we make no specific recommendations to take following detection or prediction due to the complex nature of flagging users, although \name can support a range of options for addressing these complex scenarios.

\descr{Potential Negative Outcomes.}
While there are many positive implications of combating malicious influence with systems like \name, there is a risk that attackers might become ``smarter'' and use different tactics to spread misleading content.
However, we attempt to capture the ``core'' of a influence campaign, which is network-level coordination.
For adversaries to evade detection (or become \emph{smarter}) in such a scenario would mean significantly reduce coordination or not coordinate at all.
While that strategy might stop detection, it would turn the accounts into basic bot or spam accounts and defeat the purpose of a campaign.
Similar to the risk of false positives discussed in responsible deployment, there are implications for preemptive predictions, where there is a risk of penalizing users before they have done anything wrong.
We emphasize that \name is explicitly not a call for punitive automation and is designed to assist moderators by narrowing the focus on users who are most likely to be problematic, also reducing the cognitive load on human reviewers.
In our results, we report F1/AUC to contextualize performance, but real-world deployment requires tuning for the platform's risk tolerance.

\section{Limitations and Future Work}
We rely on the ground truth provided to us by social media platforms (i.e., both Reddit and X). 
It is likely that a small number of users might be misclassified in the initial set or a small number of nodes might be malicious users in the full network.
Secondly, we utilize OpenAI's \textit{text-embedding-3-small} and SBERT for our LLM analysis.
Although, they are very well suited for classification tasks, with the advent of newer models, it is likely that some models might perform better for influence campaigns which we aim to explore and compare in future work. 
We also chunk the texts when calculating embeddings and average out the results, which might remove some structural nuance.

We also highlight campaign-level strategies through case studies to understand targeted influence with more granularity.
Next, while GraphSAGE adapts very well to social media graphs, it will be useful to compare the performance with other GNNs in combination with more sophisticated LLMs in future work.
Lastly, another future direction is to explore cross-platform campaigns and identify actors that target multiple platforms and how the affordances of the platform impacts their strategy.
We also envision future efforts to address cross-platform detection and the unique obstacles for transfer learning in troll detection systems, including adaptability across different platforms.


\descr{Acknowledgments.}
The authors would like to thank Gianluca Stringhini for data provision. 
This work was supported in part by National Science Foundation grant 212720.

\small
\bibliographystyle{apalike}
\bibliography{refs}

\end{document}